\documentclass[useAMS,usenatbib]{mn2e}

\usepackage{graphicx}

\title{On the origin of short GRBs with Extended Emission and long GRBs without associated SN}
\author[M.H.P.M. van Putten, G.M. Lee, M. Della Valle, L. Amati and A. Levinson]
{
Maurice H.P.M. van Putten$^1$ \thanks{E-mail: mvp@sejong.ac.kr}, Gyeong Min Lee$^{1}$, 
Massimo Della Valle$^{2,4}$, \newauthor\mbox{}\hskip0.08in Lorenzo Amati$^{3,4}$ and Amir Levinson$^{5}$\\
$^{1}$Astronomy and Space Science, Sejong University, 98 Gunja-Dong Gwangin-gu, Seoul 143-747, Korea\\
$^{2}$ Istituto Nazionale di Astrofisica, Osservatorio Astronomico di Capodimonte, Salita Moiariello 16, I-80131 Napoli, Italy\\
$^3$ Istituto Nazionale di Astrofisica - IASF Bologna, via P. Gobetti 101, I-40129 Bologna, Italy\\
$^4$ International Center for Relativistic Astrophysics, Piazzale della Repubblica 2, I-65122, Pescara, Italy\\
$^5$ School of Physics and Astronomy, Tel Aviv University, 69978 Tel Aviv, Israel}

\begin{document}

\date{}

\pagerange{\pageref{firstpage}--\pageref{lastpage}} \pubyear{2002}

\maketitle

\label{firstpage}

\begin{abstract}
The Burst and Transient Source Experiment (BATSE) classifies cosmological gamma-ray bursts (GRBs) into short (less than 2 s) 
and long (over 2 s) events, commonly attributed to mergers of compact objects and, respectively, peculiar core-collapse supernovae.
This standard classification has recently been challenged by the {\em Swift} discovery of short GRBs 
showing Extended Emission (SGRBEE) and nearby long GRBs without an accompanying supernova
(LGRBN). Both show an initial hard pulse, characteristic of SGRBs, followed by a long duration soft tail. 
We here consider the spectral peak energy ($E_{p,i}$)-radiated energy $(E_{iso})$ correlation and the redshift distributions to probe the 
astronomical and physical origin of these different classes of GRBs. We consider {\em Swift} events of $15$ SGRBs, $7$ SGRBEEs, 
3 LGRBNs and 230 LGRBs. 
The spectral-energy properties of the initial pulse of both SGRBEE and LGRBNs are found to coincide 
with those of SGRBs. A Monte Carlo simulation shows that the redshift distributions of SGRBs, SGRBEE and LGRBNs fall 
outside the distribution of LGRBs at 4.75$\,\sigma$, 4.67$\,\sigma$ and 4.31$\sigma$, respectively.
A distinct origin of SGRBEEs with respect to LGRBs is also supported by the elliptical host galaxies of the SGRBEE events 050509B and 050724. This combined evidence supports the hypothesis that SGRBEE and LGRBNs originate in mergers as SGRBs. Moreover,
long/soft tail of SGRB and LGRBNs satisfy the same $E_{p,i}-E_{iso}$ Amati-correlation holding for normal LGRBs. This fact points to 
rapidly rotating black holes as a common long-lived inner engine produced by different astronomical progenitors (mergers and supernovae).
\end{abstract}


\section{1. Introduction}

The bimodal distribution in the Burst and Transient Source Experiment (BATSE) catalogue of cosmological gamma-ray bursts (GRB)
reveals long GRBs (LGRB) commonly associated with core-collapse in massive stars \citep{woo93} and short GRBs (SGRBs) 
commonly associated with mergers of compact objects, i.e., neutron stars with another neutron star (NS-NS, \cite{eic89}) 
or a stellar mass black hole companion (NS-BH, \cite{pac91}). The first is supported by three pieces of evidence: $i)$ supernovae (SNe)
accompanying a few nearby events \citep{hjo11}; $ii)$ detection of SN features in the spectra of ``rebrightenings" during GRB
afterglow decay, at intermediate redshifts, most recently GRB 130427A at $z=0.34$ \citep{mel13,mas13}, up to
$z\simeq1$ \citep{del03}; $iii)$ the host galaxies are spiral and irregular with active star formation \citep{fru04} typical for environments
hosting core-collapse SNe (CC-SNe; \cite{kel08,ras08,mod11}). If detected, afterglow emissions of SGRBs tend to be very weak compared to those of LGRBs, consistent with less energy output and pointing to hosts lacking star formation. Weak X-ray afterglow 
emissions discovered by High Energy Transient Explorer-2 (HETE II) in GRB 05059B and {\em Swift} in GRB 050507 \citep{ber06a} 
were anticipated for GRBs from rotating black holes \citep{van01}. 

{\em Swift} has been key to the discovery of a diversity beyond the BATSE classification. GRB 060614 ($T_{90}=102$ s)
has no detectable supernova \citep{del06,fyn06,gal06} and GRB 050724 is a SGRB with Extended Emission (SGRBEE) with $T_{90}=69$ s in an elliptical host galaxy \citep{ber05c,bar07}, neither which is readily associated with a massive star. Since then, the list SGRBEEs has grown considerably. 

A primary question is whether the observed $z-$distribution of SGRBs is genuinely different from that of LGRBs.
If SGRBs are relatively nearby, as suggested by the currently available redshift data \citep{cow13,sie14}, here shown in Table 1, 
they may originate delayed to the cosmic star formation rate. 
In particular, a binary origin can account for an apparent offset in mean redshift of order unity \citep{por98,gue05}.
Though highly plausible \citep{nar01,fox06}, direct evidence for an origin of SGRBs in mergers is relatively weak 
compared to the evidence for a major fraction of LGRBs originating in CC-SNe. In essence,
this fact points to an intrinsic faintness of candidate progenitors. 

Here, we set out to rigorously determine whether the observed redshift distributions of SGRBs and LGRBs are distinct.  
To circumvent the limitation of small sample sizes in a model independent approach, we apply a Monte Carlo (MC) simulation to the 
extraction of samples of size 15 (for SGRBs), 7 (for SGRBEEs) and 3 (for low-$z$ LGRBs with no supernova, LGRBNs)
from the $z-$distribution of LGRBs. It obtains probabilities for mean redshifts to occur by chance
with equivalent levels of confidence for SGRBs, SGRBEEs and LGRBNs to be distinct from LGRBs.   
 
\section{MC analysis}

Table 1 lists {\em Swift} events of SGRB, SGRBEE and LGRBNs with redshifts (Fig. 1) and, when available, hosts and location in the
$E_{p,i}-E_{iso}$ plane. Note that GRB 060614 is listed both as a SGRBEE and LGRBN. 
The peak energy $E_{p,i}$ is the photon energy at which the $\nu F_\nu$ spectrum 
in the cosmological rest-frame peaks; it typically ranges from tens of keV to several hundreds of keV. $E_{iso}$ is the
isotropic-equivalent energy radiated by a GRB during its whole duration assuming spherically symmetric emission.  
$E_{iso}$ is used due to the still lacking reliable information on the degree of collimation in the individual GRB events.
In the $E_{p,i}-E_{iso}$ plane, long GRBs follow the {\em Amati-correlation} \citep{ama02,ama06}. 

In Fig. 2, both SGRBEE and LGRBN show a first spike inconsistent with the Amati-correlation for normal 
LGRBs and a long soft tail consistent with it, notably GRB 050724 \citep{bar05,ama10}, GRB060505 \citep{tho08,ofe07},
GRB060614 \citep{man14,cai09}, see further \cite{xu09}, GRB060218 \citep{ama07}, GRB071227 \citep{cai10} and GRB061021 
\citep{gol06}. This sample is key to a new diversity in GRBs, beyond the BATSE classification of SGRBs and LGRBs.

The mean values $\mu$ of the observed redshifts, i.e., $\mu_{S}^{N}$ of LGRBNs, $\mu_{EE}$ of SGRBEEs,  $\mu_{S}$ of SGRBs and 
$\mu_{L}$ of LGRBs, satisfy
\begin{eqnarray}
\mu_{S}^{N}< \mu_{EE} <\mu_{S}<\mu_{L},
\label{EQN_z}
\end{eqnarray}
where $\mu_S^{EE}=0.5286$, $\mu_S=0.8587$, $\mu_L^N=0.1870$, and $\mu_L=2.1069$.

We now consider the probability that, by mean redshift, our samples of SGRBEE ($n_1=7$), SGRB ($n_2=15$) and LGRBNs ($n_3=3$) 
are drawn from the observed distribution of LGRBs ($n=230$). Because of the small $n$ samples and the broad distribution of redshifts of LGRBs (with an observational bias towards low $z$), we proceed with an MC test by drawing samples of size $n_i$ ($i=1,2,3)$ from the distribution of the $n=230$ redshifts of the latter. Doing so $N$ times for large $N$, we obtain distributions of averages $m_i$ of the redshifts in these small $n$ samples under the Bayesian null-hypothesis of coming from the distribution of redshifts of LGRBs.
Fig. 2 shows the results for MC with $10^6$ realizations. A comparison with such MC test on a Gaussian distribution with mean $\bar{z}_L$ and standard deviation $\sigma_L=1.3459$ of LGRBs is included for reference. The results clearly demonstrate the need for a Monte Carlo simulation on the observed distribution of redshifts of LGRBs for an accurate estimate of confidence levels, especially when departures from $\bar{z}_L$ are substantial. Based on (\ref{EQN_z}), the MC analysis shows
\begin{eqnarray}
\begin{array}{rll}
\mbox{SGRBEE} & \not\subset \mbox{LGRB} &: ~~\sigma = 4.6700 \\
\mbox{SGRB}      & \not\subset \mbox{LGRB} &: ~~\sigma = 4.7530 \\
\mbox{LGRBN}   & \not\subset \mbox{LGRB} & :~~\sigma \simeq 4.3140 \\ 
\end{array}
\label{EQN_s}
\end{eqnarray}

Relative to normal LGRBs, the distinct morphology of SGRBEE and LGRBNs in Fig. 2 and redshifts (\ref{EQN_s}) provides evidence that SGRBEE and LGRBN belong to the same class of events and likely originate in mergers as do classical SGRBs. A technically long duration EE to a merger explains
LGRBNs, i.e., LGRBs without association with an SN. This interpretation is further supported by the elliptical host galaxies of SGRB 050509B and GRBEE 050724, which sets these events rigorously apart from massive star progenitors to normal LGRBs.

Our MC results (\ref{EQN_s}) also show that our sample of SGRB(EE)s has negligible contamination by short duration events derived from LGRBs with durations $T_{90}\la$ 20 s, that may be present at redshifts $z>1$ \citep{bro13}. The time scale of 20 s 
derives from a 10 s time-scale of shock break-out in CC-SNe identified in a detailed matched filtering analysis of 
the 1491 LGRBs in the BATSE catalogue \citep{van12}. Furthermore, the soft tails of both SGRBEE and LGRBN show a location 
in the $E_{p,i} - E_{iso}$ plane consistent with that of LGRBs associated with a SN and, more generally, with the Amati-correlation for normal LGRBs, that are all expected to be associated with a SN by the correlation of their redshift distribution to the cosmic star formation rate. $E_{iso}$ of the two SGRBEEs highlighted in Fig. 2 (GRB 071227 and GRB 050724) fall within the range of the E$_{iso}$'s of normal LGRBs with SNe and, in fact, are at the lower end of the $E_{iso}$ of the SGRBEEs listed in Table 1. SGRBEE and LGRBs should hereby have essentially the same observational selection effects. On this basis, (\ref{EQN_s}) shows GRBEEs and LGRBs to be distinct populations at a level of confidence exceeding 4$\,\sigma$.

\begin{table*}
\textbf{Table 1.} {\em Swift} detections of SGRB, SGRBEE$^a$ and LGRBNs sorted by redshift.\\ 
\begin{tabular}{llrrllclc}
\hline
   & & $T_{90}$ & $z$ &                   & Host$^b$                                    & $E_{iso}$$^c$ (10$^{52}$ erg)     & $E_p$$^c$ (keV)\\
\hline
SGRB  & 061201  & 0.760    &  0.111   &    	&  galaxy cluster\,[1]                              	& 0.013	 				 & 969 \\ 
	& 050509B & 0.073 	    &  0.225   &  		&  elliptical galaxy\,[2]                                  	& 0.00027$\pm$0.0001\,[3]	 & -  \\ 
	& 060502B & 0.131      &  0.287   &  		&  massive red galaxy\,[4]                           	&  0.022	  				&  193\\ 
	& 130603B & 0.18         &  0.356   &   	&  SFR\,[5]   						&  0.21$\pm$0.02\,[6]                 &  90\,[6]\\ 
	& 070724A & 0.4           &  0.457   &    	& moderate SF galaxy\,[7]                            	&   - 						&  \\ 
	& 051221A & 1.400      &  0.547   &  		&  SF, late type galaxy\,[8]                           	& 0.25\,[9] 				& \\ 
	& 131004A & 1.54         &  0.717   &   	&  low mass galaxy\,[10]                              	& -  						&   \\ 
	& 101219A & 0.6           &  0.718   &      	& faint object\,[11]                               	    	& 0.48	  				& 842 \\ 
	& 061217    & 0.210      &  0.827   &     	& faint galaxy\,[12]                                 	& 0.008\,[12] 				& \\
	& 090510    & 0.3           &  0.903   &    	& field galaxy\,[13]                       			& 3.8\,[13] \\
	& 070429B & 0.47          &  0.904  &    	&  star forming\,$\simeq$1.1$M_\odot$ yr$^{-1}$\,[14]  &  - 				& - \\ 
	& 060801    & 0.49         &  1.131   &     	&           -                              				& 0.027\,[15] \\ 
	& 100724A & 1.4            &  1.288   &  		&  probably LGRB\,[16]			       	& -  						& - \\ 
	& 050813    & 0.45         &  1.8       &  		&  galaxy cluster\,[17,18]                               & 0.017\,[18]  		 	 	& - \\
	& 090426    & 1.2           &  2.609   &       	&  irregular SF galaxy\,[19]                		& -  						& - \\
\hline
SGRBEE & 060614  $^d$ $^e$ $^f$ $^g$&   108.7 &  0.125 	&       &  faint SFR\,[20,21]   	&  0.21$\pm$0.09\,[20]     	& 55\,[20] \\
	& 050724  $^d$ $^e$ $^f$ $^g$  	&   69       &  0.258 	&       &  elliptical, weak spiral\,[22] 	&  0.0099\,[23]           & - \\ 
	& 071227A $^e$ $^f$                     	&   1.8      &  0.384  	&       &  edge-on spiral\,[24] 	&  0.008\,[25]       		& - \\ 
	& 061210  $^d$ $^e$ $^f$ $^g$  	&   85.3   &  0.41 	&       & bulge dominated\,[26]  	&  0.046\,[26]		         & - \\ 
	& 061006  $^d$ $^e$ $^f$ $^g$  	&   129.9 &  0.438 	&       & exponential disc profile\,[27]  &  0.18   			& 955 \\ 
	& 070714B $^d$ $^e$ $^f$ $^g$ 	&   64      &  0.92   	&       & moderately SF galaxy\,[28]    &  0.16\,[28,29] & -  \\ 
	& 050911  $^d$ $^e$                     	&   16.2   &  1.165  	&       & EDCC493 cluster\,[30]  	&      0.0019\,[30]  		& - \\ 
\hline
LGRBN & 060505                                    &   4     	  &  0.089   &     & spiral, ionized H, no SN\,[31] &  0.0012\,[21]\,-\,0.0039  & 120  \\ 
	& 060614  $^d$ $^e$ $^f$ $^g$  &   108.7 &  0.125   &     & faint SFR, no SN\,[20]  &  0.21$\pm$0.09\,[21]     		& - \\
	& 061021                                          &    46     & 0.3462  &     & no SN\,[32]                    & 0.68                                 		& 630\\
\hline
\label{TABE2}
\end{tabular}
\mbox{}\\
\vskip-0.15in
 $^a$ From \cite{HEA}; $^b$ galaxy type, SN association; 
$^c$ Isotropic-equivalent energy and peak energy for events with reliable estimates of the bolometric $E_{iso}$ across
a large enough energy band, under the assumption $\Omega_m=0.3$ and a Hubble constant $H_0=70$ km s$^{-1}$ Mpc$^{-1}$; 
$^d$  \citep{per09}; $^e$  \citep{nor10}; $^f$  \citep{cow12}; $^g$  \citep{gom14};
[1] \cite{ber07c} ;[2] \cite{fon10,pag06,per09} ;  [3] \cite{blo06,blo07}; [4]  \cite{blo07}; [5]  \cite{cuc13}; [6] \cite{fre13};
[7] \cite{koc10};  [8] \cite{ber05,ber07}; [9] \cite{gol05}; [10] \cite{per13}; 
[11]  \cite{per10}; [12] \cite{ber06b,deu06}; [13] \cite{rau09,gue12}; [14]  \cite{cen08}; [15] \cite{cuc06,ber07a}; 
[16]  \cite{ukw10}; 
[17] \cite{blo07,pro06,ber06a,fer07};[18] \cite{ber05a};
[19]  \cite{ant09}; 
[20] \cite{ama08}; [21] \cite{fyn06,cob06} ; 
[22]  \cite{ber05c,pag06,ber07a,fon10}; [23] \cite{pro05}; 
[24] \cite{ber07b}; [25] \cite{ber07d}; 
[26] \cite{cen06}; 
[27] \cite{fon10,ber07a}; 
[28] \cite{gra09}; [29] \cite{gra07};  
[30] \cite{ber07}; 
[31]  \cite{jak07}; [32] \cite{mor06}
\end{table*}

\begin{figure*}
\centerline{\includegraphics[width=75mm,height=50mm]{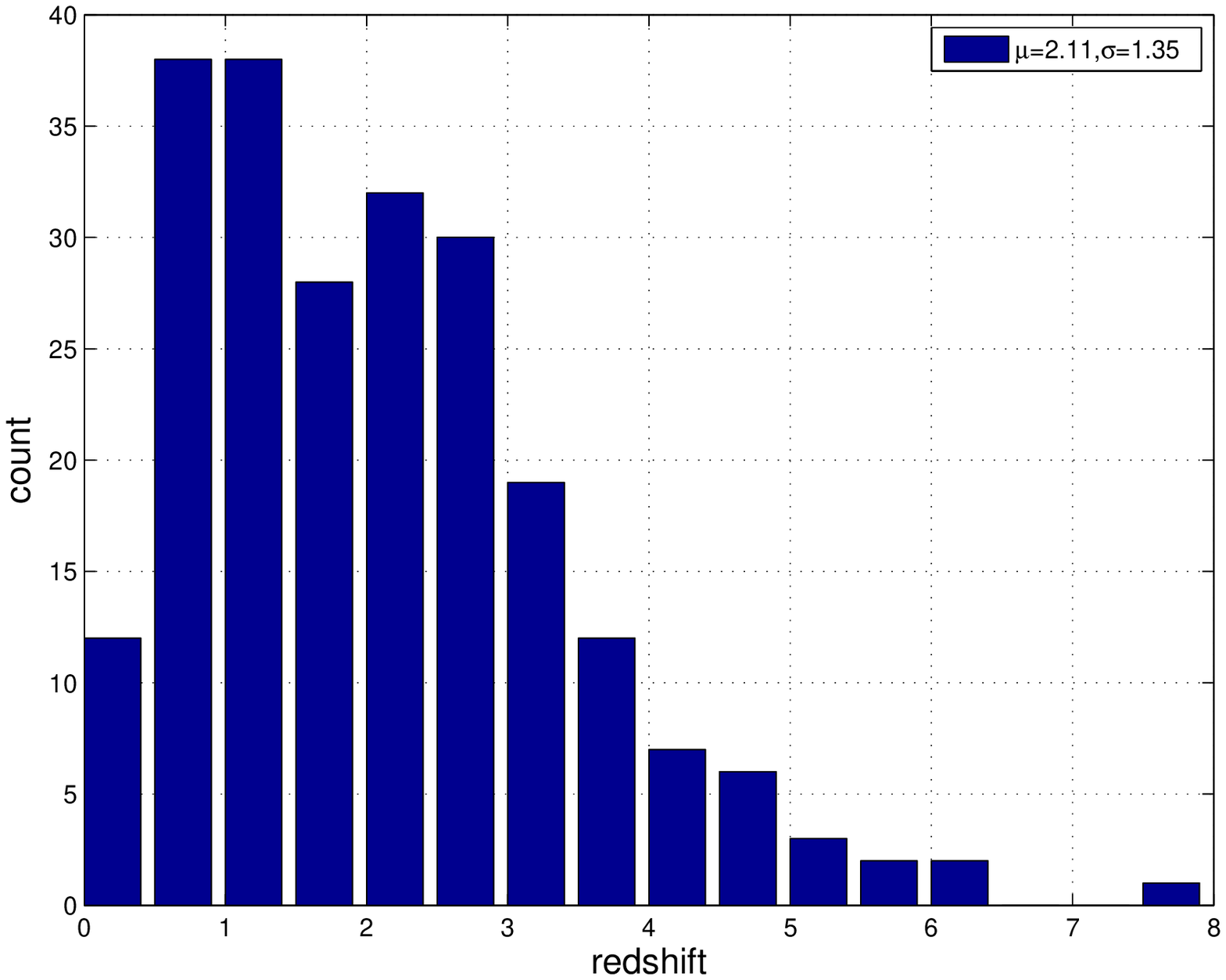}\includegraphics[width=75mm,height=50mm]{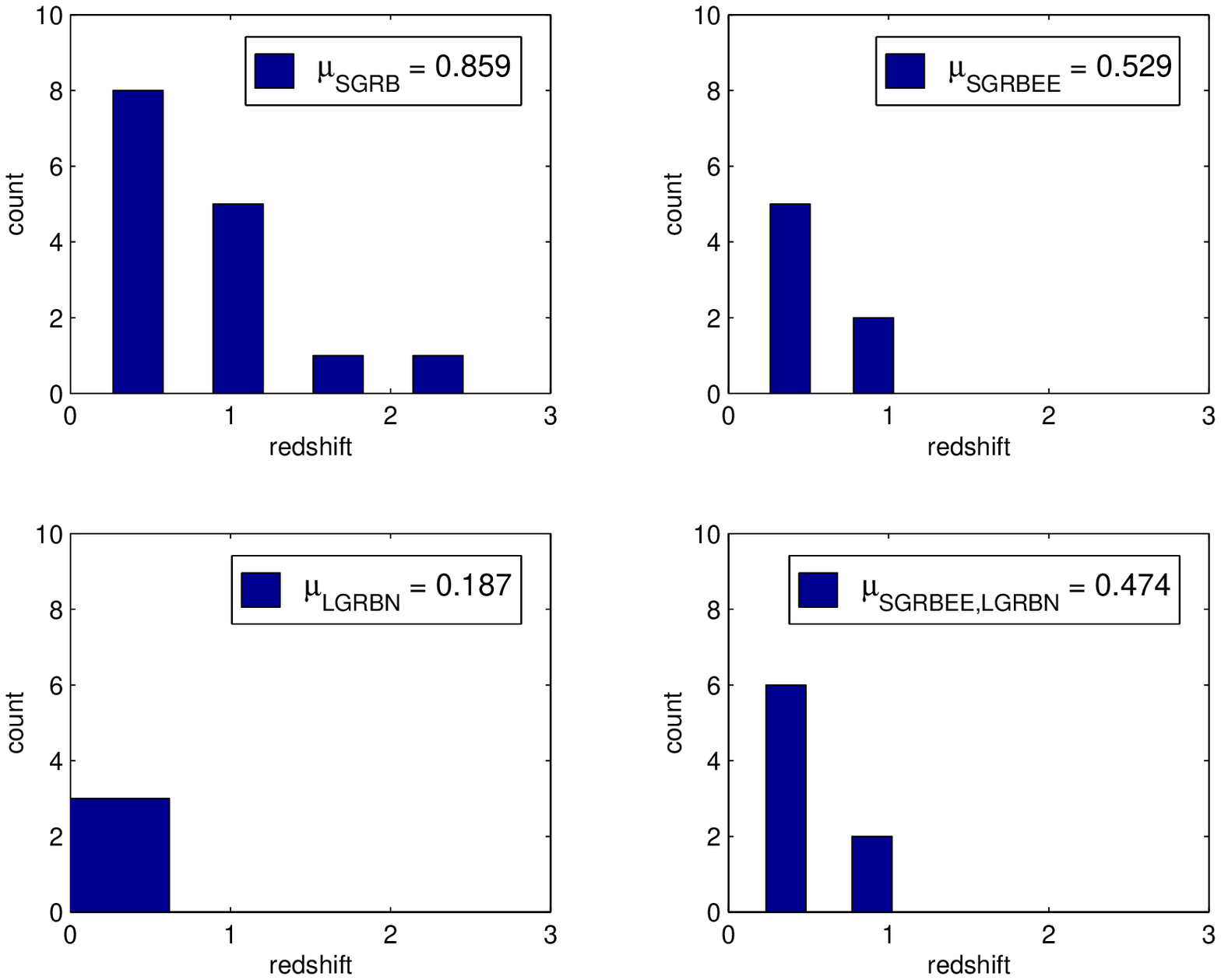}}
\caption{({Left:}) the redshifts of 230 LGRBs in the {\em Swift} catalogue shows a mean 
redshift $\mu=2.11$ with standard deviation $\sigma=1.35$. This distribution is significantly biased to 
towards low redshifts. ({Right:}) same for SGRB, SRGBEE, LGRBN and SGRBEE+LGRBNs.}
\label{fig:A1}
\end{figure*}

\begin{figure*}
\centerline{\includegraphics[width=89mm,height=76mm]{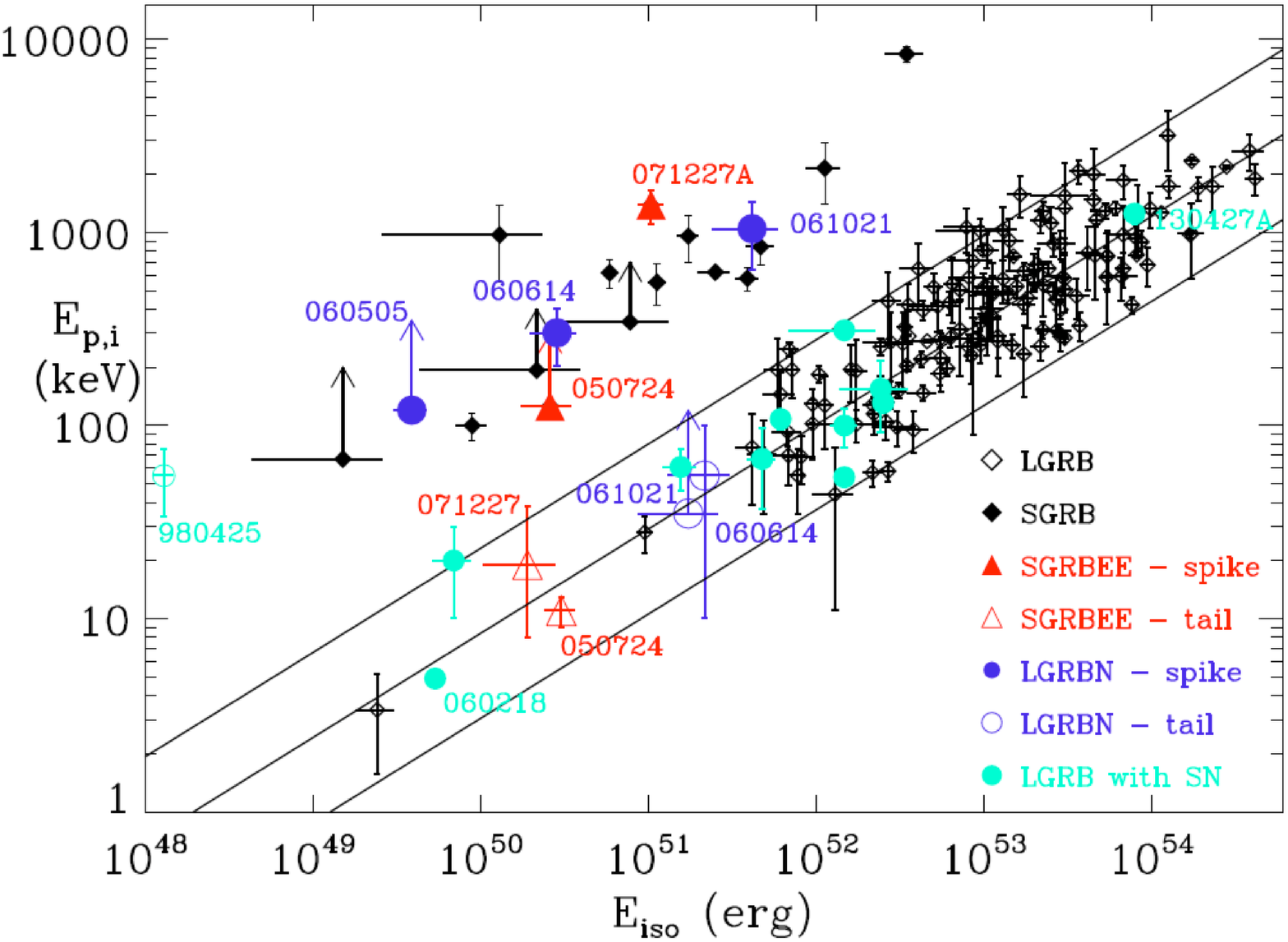}\includegraphics[width=89mm,height=75mm]{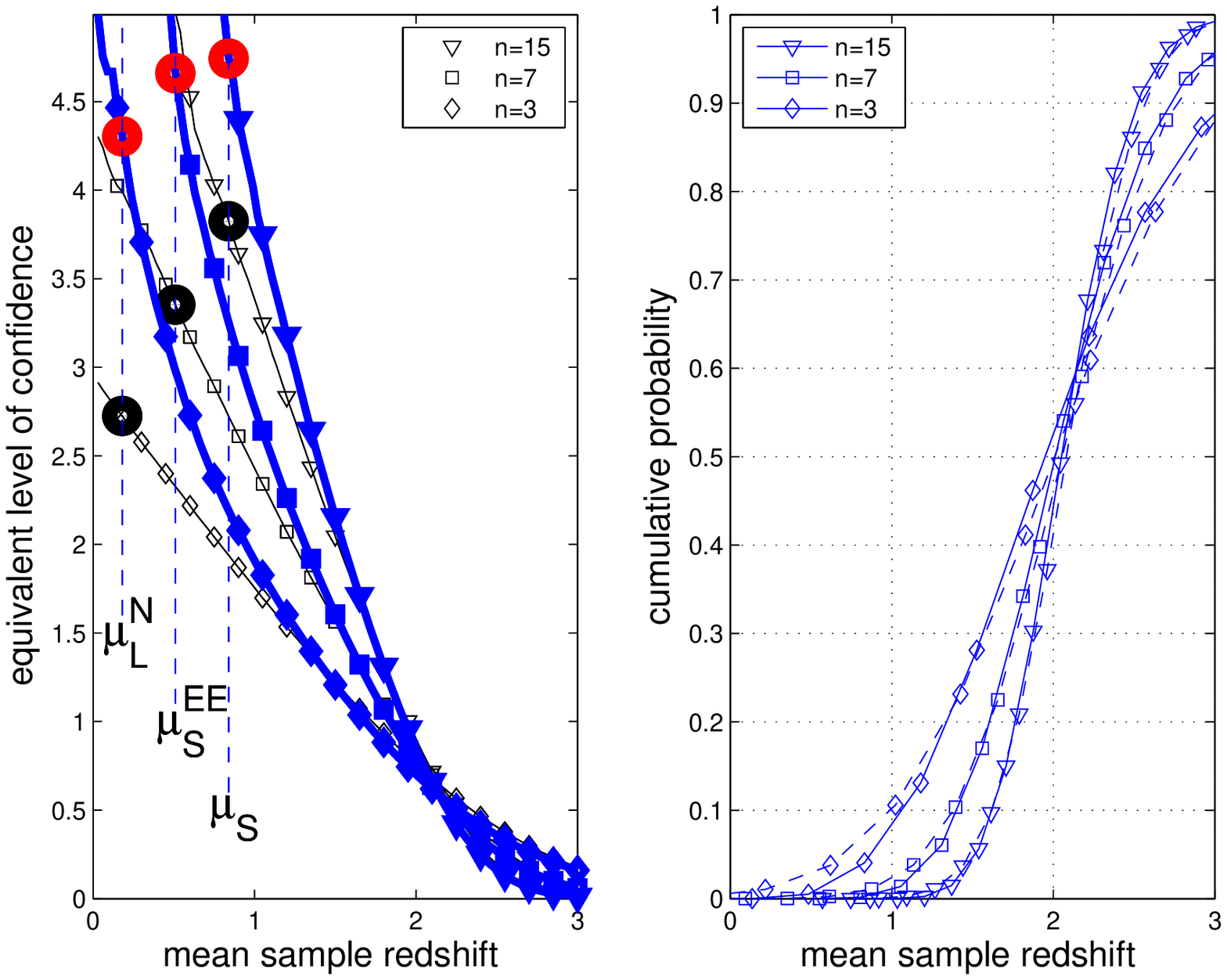}}
\caption{
({Left:}) shown are SGRBEE and LGRB(N)'s in the  $E_{p,i}-E_{iso}$ plane \citep{ama02,ama06}, including GRB-SNe
030329, 050525A, 081007,091127,100316D,101219B. The lines are the best-fit of the $E_{p,i}-E_{iso}$
correlation for normal LGRBs and its $\pm$2 $\sigma$ confidence region. The sub-energetic GRB980425/SN1998bw has a distinguished symbol. Highlighted are GRBEEs 050724 and 060614 (also a LGRBN). The tail of these two SGRBEEs (open triangles, {red}) and that of the more luminous counterparts listed in Table 1 falls well within those of LGRB with SNe indicated by medium sized filled circles ({green}). In contrast, the initial pulse of SGRBEEs (solid triangles, {red}) falls into the separate group of SGRBs, in common with the initial pulse of LGRBNs (large size filled circle, { blue}). Limits shown are 90\% confidence levels. Data mostly from \citep{ama08,ama09,can14,swi14}.
({\em Right two panels:}) an MC simulation produces probabilities $P_i(\mu<z)$ as a function of $z$ in the three
samples of size $n_i=15,7,3$ extracted, respectively, from the redshift distribution of LGRBs (blue curves). For comparison, the same is shown using a Gaussian distribution with $\sigma_L=1.35$ for the latter (black curves). In the Bayesian interpretation, $P_i(\mu<z)$ are probabilities of observing averages $\mu<z$ under the null-hypothesis of a redshift distribution given by that of LGRBs. Being small, we convert $P_i(\mu<z)$'s to equivalent confidence levels for the null-hypothesis to be false, i.e., for the three small samples to be drawn from a redshift distribution different from that of LGRBs. Vertical dashed lines show the $\mu_L$ in (\ref{EQN_z}) and their intersections with the MC results (red and black dots). Also shown are the cumulative distributions of the distributions of mean redshifts of the samples of size $n_i=15,7,3$ extracted from the redshifts of LGRBs (continuous lines), including that of a Gaussian distribution (dashed lines) with the same mean and standard deviation shown in Fig. 1.}
\label{fig:A3}
\end{figure*}

\section{Mergers and extended emission} 

A merger origin of SGRBs and SGRBEEs naturally accounts for a delay in redshift relative to the cosmic star formation rate and a 
dissociation to host galaxy type. Mergers of interest are NS-NS and NS-BH coalescence, producing stellar black holes with an 
accretion disc or torus (BHS) in common with core-collapse of high mass stars (more massive than those producing neutron stars).

A BHS may produce a long-lived inner engine in angular momentum loss from rapidly rotating black holes, as opposed to short-lived 
activity from hyper-accretion onto a slowly spinning black hole following tidal break-up of a PNS \citep{van01}. In particular, high 
resolution numerical simulations show NS-NS producing rapidly rotating black holes \citep{bai08}, while NS-BH binaries are expected 
to be diverse harbouring slowly to rapidly spinning black holes. Accordingly, we identify SGRBEE and LGRBNs with mergers involving 
the latter.

Since the soft tail of SGRBEEs and LGRBs feature share the same Amati-correlation, they likely share the same inner engine, i.e., 
rapidly rotating black holes, not PNS. (If produced in mergers, PNS are short-lived with no bearing on extended emission.) 
This safely accounts for the most hyper-energetic GRB-SNe, whose output exceeds the maximal spin energy $E_c\simeq 3\times 10^{52}$ erg of a PNS with a mass of $M=1.45M_\odot$ and radius $R=12$ km to be compared with the maximal spin energy $E_{rot}=6\times 10^{54}$ erg of a Kerr black hole of mass $M=10M_\odot$ \citep{van11b}. 

\section{Discussion}

The {\em Swift} discovery of SGRBEEs, e.g., GRB 060724, and LGRBNs, e.g., GRB 060614, highlights a new diversity in LGRBs with 
no apparent association with massive stars. 

Based on an MC analysis, the observed redshift distributions of SGRB and SGRBEEs are distinct from that of 
LGRBs at $4.75\sigma$ and $4.67\sigma$,  respectively. Further supported by their distinct morphology in the 
$E_{p,i}-E_{iso}$ 
plane and hosts that include elliptical galaxies, this points to a common origin in mergers of both, i.e., of the NS-NS and 
NS-BH variety with long-lived and short-lived inner engines, respectively.

In a common origin of SGRBEEs and LGRBNs, we identify the initial pulse with 
the ``switch on" of SGRBEEs in binary coalescence similar as in classical SGRBs (see also \cite{ghi11})
and the extended emission with a rotating black hole, rather than a PNS, slowly losing angular momentum. 

The fact that the soft tail of SGRBEE and LGRBNs follows the Amati-correlation holding for normal LGRBs points to a common long-lived inner engine which is a rotating black hole to all three groups. This is consistent with (a) the absence of any signature of PNS formation in a high-frequency analysis of {\em BeppoSAX} light curves \citep{van14} and (b) evidence for black hole spindown observed in normalized light curves in the BATSE catalogue \citep{van12}. Apart from the low number counts, the only reservation would be extremely subluminous CC-SNe (cf. \cite{pas07}), that would be undetectable in our sample LGRBNs.

Following \cite{gue07}, local rates of GRB-SNe and LGRBNs can be estimated from the observed events taking into account 
finite angular and temporal sky coverage and sensitivity distance. With no correction for beaming, we find, respectively, the
event rates 1.13$^{+1.98}_{-0.85}$  (cf. \cite{gue11}) and $0.053^{+0.10}_{-0.036}$ Gpc$^{-3}$ yr$^{-1}$.
It shows a ratio of LGRBN to GRB-SN of about 5\%, and likely no larger than about 30\%. 

\mbox{}\\
{\bf Acknowledgments.} The authors gratefully acknowledge C. Tout for comments which have considerably improved the
readability of this letter.

\label{lastpage}

\end{document}